# Correlated Charge Density Wave Insulators in Chirally Twisted Triple Bilayer Graphene

Wenxuan Wang[1†], Gengdong Zhou[1†], Wenlu Lin[1†], Zuo Feng[2], Yijie Wang[1], Miao Liang[3], Zaizhe Zhang[1], Min Wu[1], Le Liu[4], Kenji Watanabe[5], Takashi Taniguchi[6], Wei Yang[4], Guangyu Zhang[4], Kaihui Liu[2], Jinhua Gao[3], Yang Liu[1,7]\*, X.C. Xie[1,8,9], Zhida Song[1,7]\* and Xiaobo Lu[1,7]\*

[1]International Center for Quantum Materials, School of Physics, Peking University, Beijing 100871, China
[2]State Key Laboratory for Mesoscopic Physics, Frontiers Science Centre for Nano-optoelectronics, School of Physics, Peking University, 100871, Beijing, China
[3]School of Physics and Wuhan National High Magnetic Field Center, Huazhong University of Science and Technology, Wuhan 430074, China
[4]Beijing National Laboratory for Condensed Matter Physics, Institute of Physics, Chinese Academy of Sciences, 100190, Beijing, China
[5]Research Center for Electronic and Optical Materials, National Institute of Material Sciences, 1-1 Namiki, Tsukuba 305-0044, Japan
[6]Research Center for Materials Nanoarchitectonics, National Institute of Material Sciences, 1-1 Namiki, Tsukuba 305-0044, Japan
[7]Collaborative Innovation Center of Quantum Matter, Beijing 100871, China
[8]Institute for Nanoelectronic Devices and Quantum Computing, Fudan University, Shanghai 200433, China
[9]Hefei National Laboratory, Hefei 230088, China

†W.W., D.G., and W.L. contributed equally to this work.
\*E-mail: xiaobolu@pku.edu.cn; songzd@pku.edu.cn; Liuyang02@pku.edu.cn

**Electrons residing in flat-band system can play a vital role in triggering spectacular phenomenology due to relatively large interactions and spontaneous breaking of different degeneracies. In this work we demonstrate chirally twisted triple bilayer graphene, a new moiré structure formed by three pieces of helically stacked Bernal bilayer graphene, as a highly tunable flat-band system. In addition to the correlated insulators showing at integer moiré fillings, commonly attributed to interaction induced symmetry broken isospin flavors in graphene, we observe abundant insulating states at half-integer moiré fillings, suggesting a longer-range interaction and the formation of charge density wave insulators which spontaneously break the moiré translation symmetry. With weak out-of-plane magnetic field applied, as observed half-integer filling states are enhanced and more quarter-integer filling states appear, pointing towards further quadrupling moiré unit cells. The insulating states at fractional fillings combined with Hartree-Fock calculations demonstrate the observation of a new type of correlated charge density wave insulators in graphene and points to a new accessible twist manner engineering correlated moiré electronics.**

In the realm of condensed matter physics, the behavior of correlated electrons presents a fascinating playground for discovering new quantum phenomena. Particularly, when electrons interact strongly with each other, the system will exhibit a variety of quantum ground states that break different symmetries. For correlated electrons, different values of kinetic energy $t$ can drive the electronic system into different quantum phases. When $t$ approaches zero, electrons are dominated by Coulomb interaction and form charge ordered Wigner crystal states which spontaneously break the translational symmetry and are crystalized into triangle electronic lattices [1]. When $t$ is comparable with Coulomb repulsion energy $U$, the system will be driven into metallic correlated charge density wave (CCDW) states with broken translational symmetry and weaker spatial modulation of electrons. However, the CCDW insulators with gap openings at fractional fillings of the electronic band which exist in an intermediate range of correlated strength are still scarcely reported. Understanding CCDW are important to elucidate the complex collective behavior of correlated electrons since CCDW usually leads to exotic quantum phenomena. For instance, in certain materials, CCDWs coexist with or evolve into phases showing superconductivity and novel magnetism, enriching the landscape of quantum materials [2-13].

Moiré engineering of two-dimensional materials provides a highly tunable platform for studying correlated phenomena including CCDW insulators by creating the flat bands which in essence reduce the kinetic energy $t$ of electrons. Insulating states at fractional fillings of moiré bands have been observed in both transitional metal dichalcogenides and rhombohedral graphene moiré systems in ultra-strong interacting regime. These fractional filled insulating states accompany a series of remarkable findings, i.e., fractional Chern insulators [14-18] and generalized Wigner crystals [19-24]. These states mainly occur at odd denominator fractional fillings (i.e., 1/3, 2/3, 2/5 etc) and the electrons are highly localized at specific stacking sites. In twisted mono-bilayer graphene [11] and twisted tri-layer graphene [12], metallic CCDW modulation has been observed at even denominator fractional fillings. CCDW states exhibiting non-trivial topology [11] and their interplay with superconductivity [12] have been reported, these states are mostly metallic nevertheless due to weaker interactions to form an insulator compared with transitional metal dichalcogenides and rhombohedral graphene systems.

In this work, we report the observation of even-denominator fractional CCDW insulators in chirally twisted triple bilayer graphene (CTTBG) which has been predicted to be a flat-band system [26]. As shown in Fig.1a, the moiré system consists of three pieces of Bernal stacked bilayer graphene chirally stacked on top of each other with a same rotation angle $\theta$. Fig.1b shows the Brillouin zones of CTTBG. In sharp contrast to alternately twisted trilayer moiré systems with mirror symmetry which host only one well defined moiré Brillouin zone, CTTBG have two equal-sized moiré Brillouin zones with a misorientation of $\theta$ generated from adjacent two layers, which is similar with previously reported helical trilayer graphene [27-28]. In real space associated with the two Brillouin zones, there are two sets of moiré superlattices which have the same periodicity $\lambda_m \sim a/\theta$ with small-angle approximation and a rotation angle $\theta$ with each other (a is the lattice constant of Bernal bilayer graphene). The two sets of misaligned moiré superlattices further generate a super moiré structure with a periodicity of $\lambda_{mm} \sim \lambda_m /\theta$ which is more than one orders larger than $\lambda_m$ at small angle limit $\theta <2°$. The super moiré structure with ultra-large periodicity ($\lambda_{mm} \sim 280$nm for $\theta \sim 1.7°$) largely retain the electronic properties dominated by original lattice and the first order moiré for the carrier density in the experimentally accessible range ($n \sim 10^{10}$-$10^{13}$cm$^{-2}$) whereas the super moiré with $\theta <2°$ mainly affect electrons with lower energy which are usually dominated by disorder. Fig. 1e shows the single-particle band structure of CTTBG ($\theta \sim 1.7°$) calculated with a continuum model at zero perpendicular electric displacement field $D$. Similar with magic-angle twisted bilayer graphene [29-31], CTTBG host a pair of flat moiré bands well isolated from other dispersive bands at higher energy. As shown in Fig. 1e, when a finite $D$ is applied, the intertwined flat valance band and conduction band separate, opening up a band gap at charge neutrality. The ultra-flat moiré bands (band width, $W<10$meV) with high tunability can exist for a relatively wide range of θ (Fig.1c), making CTTBG a new platform for exploring interacting physics.

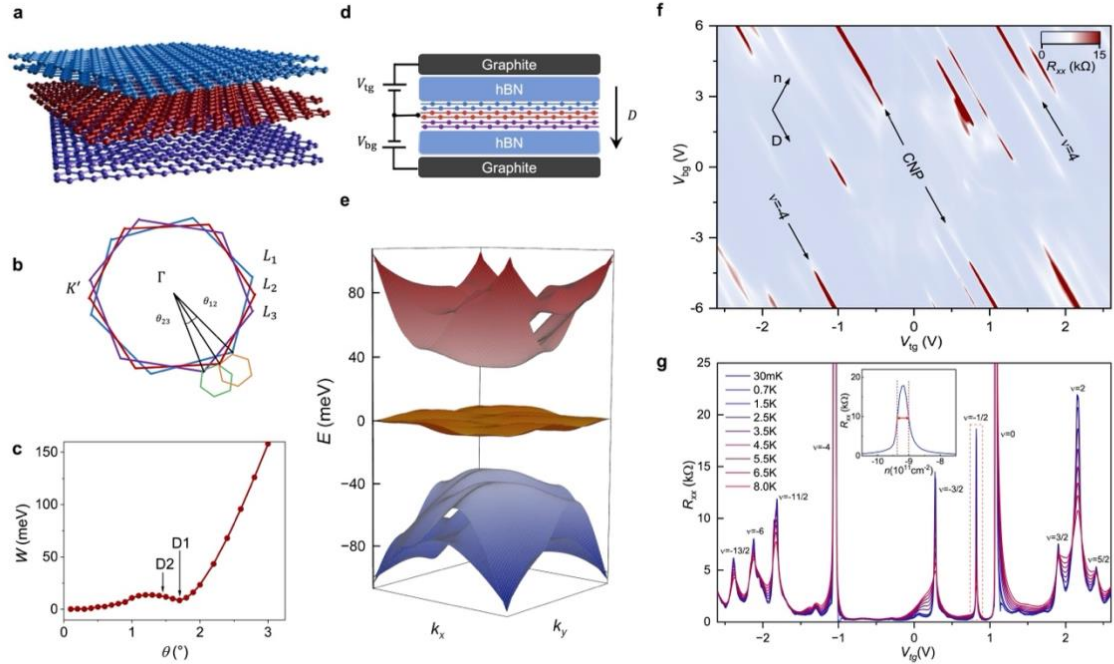

**Fig.1| Abundance of insulating states in CTTBG. a,** CTTBG consists of three layers of Bernal bilayer graphene rotated in the same direction by the same twist angle $\theta\sim1.7°$. **b,** The Brillouin zone of CTTBG. **c,** Band width of central flat band as a function of twist angle. The arrows denote the twist angle of device D1 (1.7°) and device D2 (1.46°) **d,** Schematic of our CTTBG samples which are encapsuled by hBN and gated by top and bottom graphite layers. The arrows indicate the direction of displacement field $D$ and carrier density $n$. **e,** Band structure of CTTBG at $\theta=1.7°$ achieved by non-interacting model. **F,** Longitudinal resistance $R_{xx}$ of device D1 as a function of top gate voltage $V_{tg}$ and bottom gate voltage $V_{bg}$ at $T=30$mK. **g,** Longitudinal resistance $R_{xx}$ dependence for device D1 on function of top gate voltage $V_{tg}$ at bottom gate voltage $V_{bg}=-6$V at temperature ranging from 30mK to 9K. The inset shows the zoom-in details of $R_{xx}$ as a function of carrier density for one of the peaks indicated by the red rectangle with the red dashed line representing the full width at half maximum $\Delta n\sim4\times10^{10}$cm$^{-2}$.

The CTTBG devices with a schematic shown in Fig. 1d are fabricated by a standard cut-and-stack technique [32-34] (see SI for details) with top and bottom graphite gates. We measured two CTTBG devices with twist angle $\theta = 1.75°$ (device D1) and $\theta = 1.46°$ (device D2). More geometrical details of the two devices are shown in Fig. S3. Two graphite gates enable us to independently control carrier density $n = (C_{tg}V_{tg} + C_{bg}V_{bg})/e$ and electric displacement field $D = (C_{tg}V_{tg} − C_{bg}V_{bg})/2\varepsilon_0$. Here $C_{tg}$ ($C_{bg}$) is the geometrical capacitance between top (bottom) graphite gate and CTTBG (see SIs for details), $V_{tg}$ ($V_{bg}$) is applied top (bottom) gate voltage and $e$ is the elementary charge. Fig. 1f shows the color plot of longitudinal resistance $R_{xx}$ measured in device D1 as a function of $V_{tg}$ and $V_{bg}$ at estimated base electronic temperature $T\sim30$mK. To make accurate assignment of the filling factors for the flat bands, we have performed magneto-transport measurements which can give the accurate value for $C_{tg}$ and $C_{bg}$ (Fig. S4-5). The two resistive states indicated with arrows in Fig. 1f originate from the single-particle gaps between the fourfold spin-valley degenerated flat bands and the nearest dispersive bands, corresponding to four electrons ($v=4$, $n=7.19\times10^{12}$cm$^{-2}$) and four holes ($v=-4$, $n=-6.98\times10^{12}$cm$^{-2}$) filled in a moiré unit cell. We notice that there is a slight difference between the density magnitudes for the filling factors of $v=-4$ and $v=4$. This is due to slightly electron doped charge neutral point (CNP) locating at $n\sim9.4\times10^{10}$/cm$^2$ for device D1 and $n\sim8.5\times10^{10}$/cm$^2$ for device D2. Beyond the scope of non-interacting bands, we have observed multiple resistive peaks as well as gap opening at charge neutrality ($v=0$) at finite $D$. Fig. 1g illustrates the measured data of $R_{xx}$ as a function of $V_{tg}$ at different temperature and $V_{bg}=-6$V. In addition to the most two resistive peaks corresponding to $v=-4$ and $v=0$, all resistive states show thermally active behavior (Fig.1g) and sign reversal in Hall resistance (Fig. S6), indicating opening of correlated gaps. We also note that some states exhibit very sharp resistance peaks. The inset of Fig. 1g shows the zoom-in details for one of the states with the value of full width at half maximum in terms of carrier density as low as $\Delta n \sim 4\times10^{10}$cm$^{-2}$, indicating the ultra-low inhomogeneity in the device.

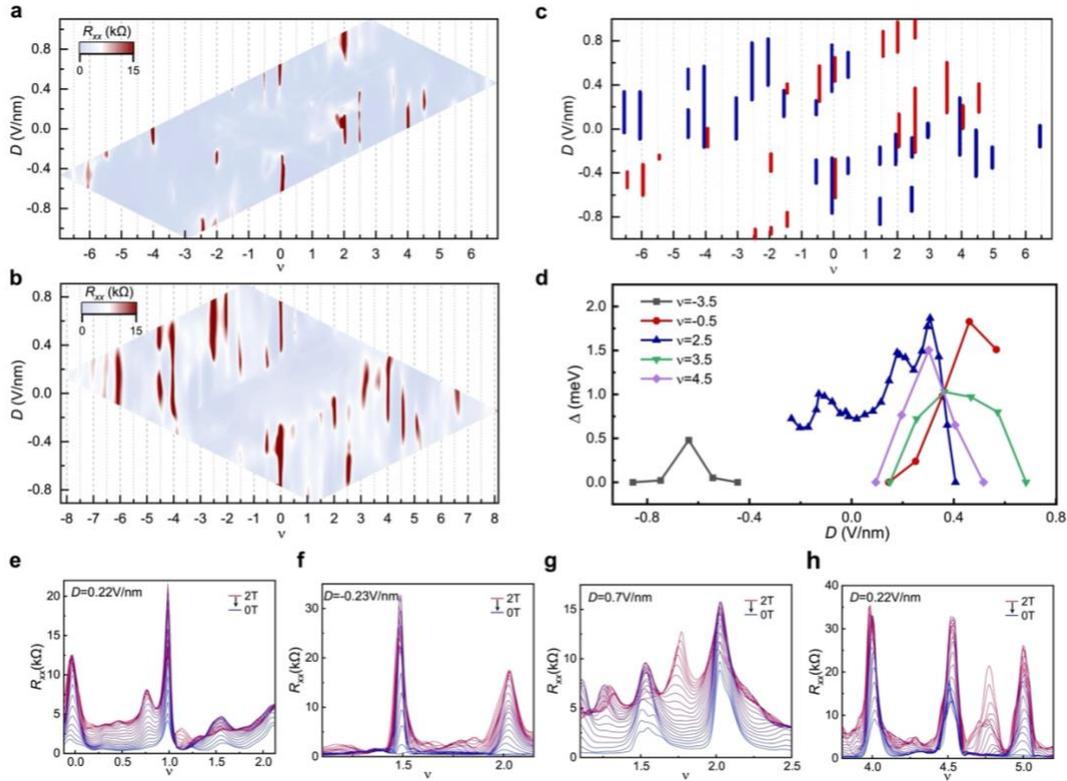

**Fig.2| Robust CCDW insulating states. a-b,** Longitudinal resistance $R_{xx}$ as a function of filling factor $v$ and displacement field $D$ for device D1 **(a)** and device D2 **(b)**. **c,** Summarized insulating states in device D1 (red lines) and device D2 (blue lines). **D,** Gap values of different half-integer

filled states versus displacement field $D$ for device D1. **E-h,** Longitudinal resistance $R_{xx}$ as a function of filling factor $\nu$ at perpendicular magnetic field $B_\perp$ varying from 0T to 2T at optimal displacement field $D$.

The dependence of $R_{xx}$ on the displacement field $D$ and moiré filling factor $\nu$ is shown in Fig. 2a. Interestingly, all observed resistive states fall into integer and half- integer moiré fillings. As shown in the $\nu$-$D$ phase diagram, the $\nu = \pm 4$ states and charge neutral point show gap opening behavior at finite displacement field. Both $\nu = \pm 2$ states exhibit notable dependance on $D$, with a gap size of ~3.85 meV at D~0.03V/nm for $\nu=2$ (Fig.S9). We also note that the gap of $\nu=2$ state first closes around $D=0.15$V/nm and reopens at higher $D$ which possibly indicates layer polarization at higher $D$. The gap opening behavior at $\nu = \pm 4$ states and charge neutrality is consistent with the non-interacting band structure in Fig.S1. Similar with other flat band systems in graphene, the occurrence of $\nu = \pm 2$ states is mostly attributed to the breaking of the fourfold spin-valley degeneracy due to interaction. Beyond the fourfold spin-valley flavors, we have further observed multiple insulating states at half-integer fillings, indicating spontaneous breaking of the translational symmetry and the formation of CCDW states. The observed half-integer fillings states exhibit clear gap opening evidenced by thermally activated behavior of $R_{xx}$ and sign reversal of Hall resistance $R_{xy}$ (Fig.S6), which is in sharp contrast with previous observed compressible CDW states in other graphene systems [10-11]. Similar with $\nu = \pm 2$ states, the half-integer filled CCDW states show strongly non-monotonic dependence on $D$ (Fig. 2d), which most likely originates from the deformation of the non-interaction band at different $D$. To rule out the possibility that as observed half-integer filling states happen to be artifacts induced by super moiré due to nonidentical twist angles in adjacent two pairs of bilayer graphene [35-39], we have fabricated a second device with slightly different twist angle θ=1.46°, which hosts similar flat bands (Fig. S1). Phenomenologically, most of the features observed in device D1 are reproduced (Fig. 2c) with more details shown in Fig. S10-12. Fig. 2c schematically displays all cumulative states observed in two different devices. The cascaded insulating states stabilizing at both integer and half-integer moiré filling strongly suggest CTTBG is a unique strongly correlated system simultaneously breaking isospin degeneracy and translational symmetry.

Interestingly, both as observed integer filled correlated states and half-integer filled CCDW states can be further enhanced by weak perpendicular magnetic field $B_\perp$. $R_{xx}$ as a function of moiré filling factor $\nu$ at fixed $D$ and different $B_\perp$ are shown in Fig. 2e-h. As the $B_\perp$ field is increased, the resistivity of the original states is enhanced ($\nu=0, 1, 4, 9/2$) and moreover new integer and half-integer filled states emerge ($\nu=3/2, 2, 5$). As the $B_\perp$ field is further increased but still before clear Landau levels involved, we have observed more resistive states appear at quarter-integer fillings (i.e., $\nu=3/4, 5/4, 7/4, 19/4$ in device D1). Device2 exhibits similar behavior (Fig. S11). The evolution of all resistive states with $B_\perp$ is further displayed in Fig. S12-13. These new appeared quarter-integer filling states also show thermally activated behavior for $R_{xx}$, indicating the formation of new correlated gaps and further quadrupling of unit cell defined by original moiré. (Fig. S14). We suggest that the enhancement of both integer and fractional filling correlated states under $B_\perp$ field in CCTBG can be attributed to the suppression of kinetic energy by $B_\perp$ field. This suppression allows the interaction energy to play a more dominant role, consequently intensifying the correlation effect. This is a reminiscent of the magnetic-field-induced Wigner crystal, wherein a $B_\perp$ field turns electrons into more localized cyclotron orbits and Coulomb repulsion organizes them into lattice subsequently [40]. This is in sharp contrast with magic angle twisted bilayer graphene, wherein the electrons are already highly localized at correlated states (Fig. S18) and $B_\perp$ field plays a less important role [30, 31].

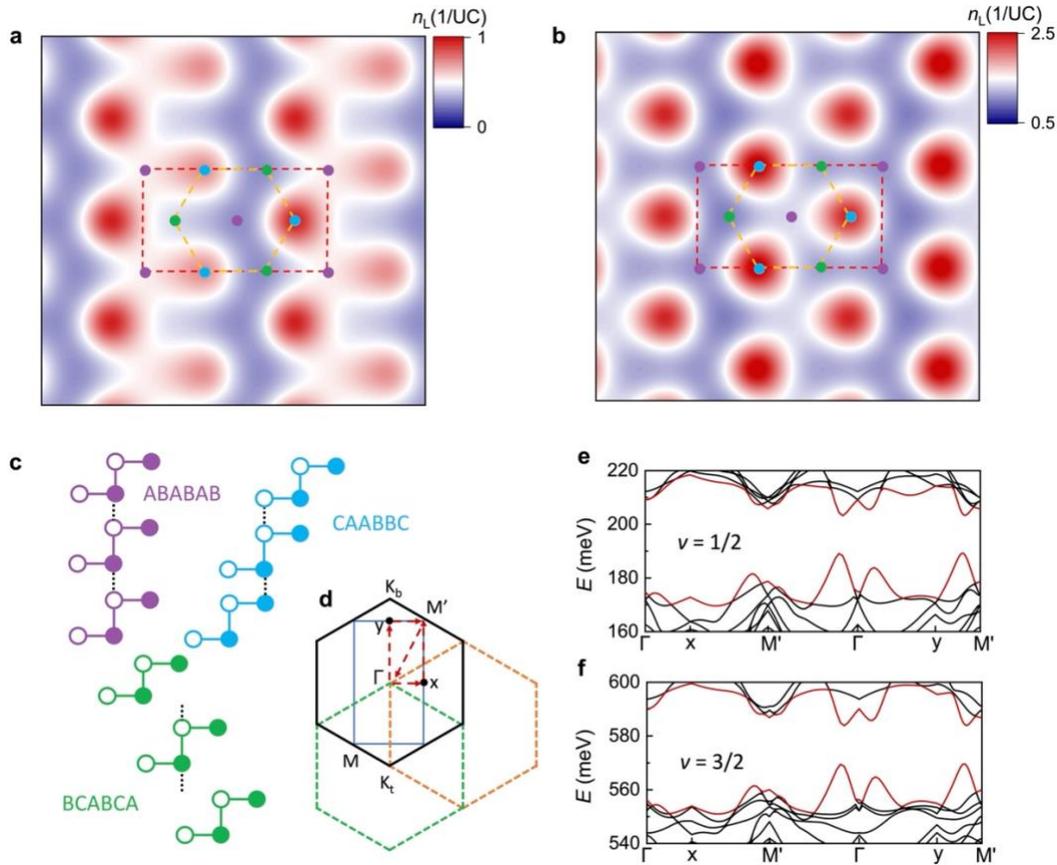

**Fig.3| Electron distributions at half-integer filled CCDW states. a-b,** Electron distribution by Hartree-Fock calculation in real space for $v=1/2$ **(a)** and $v=3/2$ **(b)** for ABABAB stacking order. The value $n_L$ represent for the number of electrons per moiré unit cell. The circle dots represent for different stacking sites. Purple dots represent for ABABAB stacking sites, blue dots represent for CAABBC stacking sites, green dots represent for BCABCA stacking sites. The orange dashed hexagons represents for the original moiré unit cell, the red dashed rectangles represent for the expanded moiré unit cell formed by CDW. **C,** Sketch of different stacking sites for ABABAB stacking order. **d,** The projection path of Hartree-Fock band. The red dashed arrows represent for the projection path. The green and orange dashed hexagons represent for the Brillouin zone in Fig.1b. **e-f,** Band structure of $v=1/2$ **(e)** and $v=3/2$ **(f)** achieved by Hartree-Fock calculation.

To reveal the underlying mechanism of the CCDW states in CTTBG, we have performed Hartree-Fock calculations. Generally, CTTBG has two kinds of stacking orders, distinguished by how the moiré patterns formed by the top and middle bilayers aligns with the one formed by the middle and the bottom bilayers. Each moiré pattern contains ABAB, ABBC and ABCA stacking regions, and the ABAB stacking region in the top moiré pattern can align with all these three types of regions in the bottom one. However, the latter two stacking orders can be transformed to each other by $C_{2x}$ rotation, so only two kinds of stacking orders are inequivalent, which we denote as ABABAB and ABABBC. (Fig. S2). The first stacking order hosts a $C_{2x}$ symmetry which can lead to symmetric phase diagrams in Fig. 2a and 2c in terms of electric field $D$ but both devices exhibit asymmetric behavior. However, this $C_{2x}$ symmetry can be easily broken by strain or twist angle inhomogeneity. The second stacking order does not host $C_{2x}$ symmetry but the corresponding central band width is very large compared to the first stacking order. Unlike the bandwidth minimum near θ=1.7° (Fig.1c) for the first stacking order, the bandwidth of the second stacking order monotonically increases with twist angle (Fig. S1), which contradicts to the abundance of correlated states we observed at device D1, highly indicating the validity of the first stacking order which is adopted in our calculations with the results shown in Fig.3. As shown in Fig. 3a, the electrons at $v$=1/2 states redistribute on original moiré background, forming a stripe phase, with electron density maximized on CAABBC site and minimized at ABABAB. Interestingly, the charge density fluctuation can even induce hole occupied site at $v$=1/2 state. The emergence of hole filled pockets can be qualitatively understood as a result of layer polarization at finite $D$ where the holes locate at a different layer from electrons with the overall electron density kept at the value corresponding to 1/2 electron per moiré unit cell. Similar density waves have been captures at $v$=3/2 (Fig. 3b) and quarter filled states (Fig. S16). Though ABABBC stacking order is not valid for our experimental result, it can also form CDW order due to our calculation (Fig. S17). According to our calculation, the electrons at different fractional fillings are all widely distributed across the whole moiré unit cell with a soft commensurate modulations of charge density, forming CCDW insulators, which is in sharp contrast with magic angle bilayer graphene (Fig. S18). In consistent with our experimental results, Hartree-Fock calculations demonstrate correlated gaps at half-integer fillings as shown in Fig. 3d-f. To further clarify the origin of CCDW states, we calculate the electron real space density profiles of the central flat bands (Fig.18). Compared with magic angle twist bilayer graphene, the density profiles of the flat bands in CTTBG are more extended. This means that nonlocal Coulomb interactions are more important in CTTBG, which can intrigue the existence of CCDW as inter-site repulsion disfavors electrons occupying neighbor sites.

In conclusion, we have observed CCDW insulators by creating a new type of flat-band moiré system-CCTBG with twist angles $\theta$ ~1.7°. The CCDW insulating states are clearly observed at half-integer moiré fillings (i.e. 1/2, 3/2, 5/2 and 7/2 etc) at zero $B_\perp$ field and quarter-integer filling states (i.e. 1/4, -7/4 and -9/4 etc) at weak $B_\perp$ field. Our experimental results combined with Hartree-Fock calculations reveal spontaneous translational symmetry breaking and the formation of long-range charge orders beyond the scale of moiré unit cell. Our findings directly place CTTBG a special correlated moiré system in which the relative correlation strength at fractional moiré fillings locates in a unique range and possible novel correlated phases leave for further investigations.

We are grateful for fruitful discussions with Haiwen Liu and Ming Lu. X.L. acknowledges support from the National Key R&D Program (Grant nos. 2022YFA1403502) and the National Natural Science Foundation of China (Grant Nos. 12274006 and 12141401). X.X. acknowledges support from Innovation Program for Quantum Science and Technology (Grant no. 2021ZD0302400). K.W. and T.T. acknowledge support from the JSPS KAKENHI (Grant nos. 21H05233 and 23H02052) and World Premier International Research Center Initiative (WPI), MEXT, Japan.

X.L. and W.W. conceived and designed the experiments; W.W fabricated the devices. W.W. and W.L. performed the measurement with help from Z.F., Z.Z., M.W., L.L., W.Y. and Y.L.; W.W., X.L., X.X., and Z.S. analyzed the data; GD.Z., M.L., J.G., X.X. and Z.S. performed the theoretical modeling; T.T. and K.W. contributed materials; X.L., Y.L., X.X., K.L. and GY.Z. supported the experiments: W.W., GD.Z., Z.S. and L.L. wrote the paper.

The authors declare no competing financial and non-financial interests.**References**
1. Wigner, E., On the Interaction of Electrons in Metals. *Phys. Rev.* **46**, 1002–1011 (1934).
2. Chang, J., Blackburn, E., Holmes, A. *et al.* Direct observation of competition between superconductivity and charge density wave order in YBa$_2$Cu$_3$O$_{6.67}$. *Nature Phys* **8**, 871–876 (2012).
3. Campi, G., Bianconi, A., Poccia, N. *et al.* Inhomogeneity of charge-density-wave order and quenched disorder in a high-$T_c$ superconductor. *Nature* **525**, 359–362 (2015).
4. Langer, M., Kisiel, M., Pawlak, R. *et al.* Giant frictional dissipation peaks and charge-density-wave slips at the NbSe$_2$ surface. *Nature Mater* **13**, 173–177 (2014).
5. Xi, X., Zhao, L., Wang, Z. *et al.* Strongly enhanced charge-density-wave order in monolayer NbSe$_2$. *Nature Nanotech* **10**, 765–769 (2015).
6. Teng, X., Oh, J.S., Tan, H. *et al.* Magnetism and charge density wave order in kagome FeGe. *Nat. Phys.* **19**, 814–822 (2023).
7. Teng, X., Chen, L., Ye, F. *et al.* Discovery of charge density wave in a kagome lattice antiferromagnet. *Nature* **609**, 490–495 (2022).
8. Nie, L., Sun, K., Ma, W. *et al.* Charge-density-wave-driven electronic nematicity in a kagome superconductor. *Nature* **604**, 59–64 (2022).
9. Joe, Y., Chen, X., Ghaemi, P. *et al.* Emergence of charge density wave domain walls above the superconducting dome in 1$T$-TiSe$_2$. *Nature Phys* **10**, 421–425 (2014).
10. Gruner, T., Jang, D., Huesges, Z. *et al.* Charge density wave quantum critical point with strong enhancement of superconductivity. *Nature Phys* **13**, 967–972 (2017).
11. Polshyn, H., Zhang, Y., Kumar, M.A. *et al.* Topological charge density waves at half-integer filling of a moiré superlattice. *Nat. Phys.* **18**, 42–47 (2022).
12. Siriviboon, P., Lin J., Liu, X. *et al.* A new flavor of correlation and superconductivity in small twist-angle trilayer graphene. Preprint at https://arxiv.org/abs/2112.07127v2 (2022).
13. Chen, G., Sharpe, A. L.; Fox, E. J. *et al* Tunable Orbital Ferromagnetism at Noninteger Filling of a Moiré Superlattice. *Nano Lett.* **22**, 238– 245 (2022).
14. Lu, Z., Han, T., Yao, Y. *et al.* Fractional Quantum Anomalous Hall Effect in a Graphene Moiré Superlattice. Preprint at https://arxiv.org/abs/2309.17436 (2023).
15. Cai, J., Anderson, E., Wang, C. *et al.* Signatures of fractional quantum anomalous Hall states in twisted MoTe2. Nature 622, 63–68 (2023).
16. Xu, F., Sun, Z., Jia, T. *et al.* Observation of Integer and Fractional Quantum Anomalous Hall Effects in Twisted Bilayer MoTe$_2$. *Phys.Rev.X.* 13, 031037 (2023).
17. Park, H., Cai, J., Anderson, E. *et al.* Observation of fractionally quantized anomalous Hall effect. *Nature* **622**, 74–79 (2023).
18. Zeng, Y., Xia, Z., Kang, K. *el al.* Integer and fractional Chern insulators in twisted bilayer MoTe2. Preprint at https://arxiv.org/abs/2305.00973 (2023).
19. Chen, G., Zhang, Y., Sharpe A. *et al.* Magnetic field stabilized Wigner crystal states in a graphene moiré superlattice. *Nano Lett.* **23**, 15, 7023–7028 (2023).
20. Regan, E.C., Wang, D., Jin, C. *et al.* Mott and generalized Wigner crystal states in WSe$_2$/WS$_2$ moiré superlattices. *Nature* **579**, 359–363 (2020).
21. Li, H., Li, S., Regan, E.C. *et al.* Imaging two-dimensional generalized Wigner crystals. *Nature* **597**, 650–654 (2021).

# Correlated Charge Density Wave Insulators in Chirally Twisted Triple Bilayer Graphene

Wenxuan Wang[1†], Gengdong Zhou[1†], Wenlu Lin[1†], Zuo Feng[2], Yijie Wang[1], Miao Liang[3], Zaizhe Zhang[1], Min Wu[1], Le Liu[4], Kenji Watanabe[5], Takashi Taniguchi[6], Wei Yang[4], Guangyu Zhang[4], Kaihui Liu[2], Jinhua Gao[3], Yang Liu[1,7]*, X.C. Xie[1,8,9], Zhida Song[1,7]* and Xiaobo Lu[1,7]*

[1]International Center for Quantum Materials, School of Physics, Peking University, Beijing 100871, China
[2]State Key Laboratory for Mesoscopic Physics, Frontiers Science Centre for Nano-optoelectronics, School of Physics, Peking University, 100871, Beijing, China
[3]School of Physics and Wuhan National High Magnetic Field Center, Huazhong University of Science and Technology, Wuhan 430074, China
[4]Beijing National Laboratory for Condensed Matter Physics, Institute of Physics, Chinese Academy of Sciences, 100190, Beijing, China
[5]Research Center for Electronic and Optical Materials, National Institute of Material Sciences, 1-1 Namiki, Tsukuba 305-0044, Japan
[6]Research Center for Materials Nanoarchitectonics, National Institute of Material Sciences, 1-1 Namiki, Tsukuba 305-0044, Japan
[7]Collaborative Innovation Center of Quantum Matter, Beijing 100871, China
[8]Institute for Nanoelectronic Devices and Quantum Computing, Fudan University, Shanghai 200433, China
[9]Hefei National Laboratory, Hefei 230088, China

†W.W., D.G., and W.L. contributed equally to this work.
*E-mail: xiaobolu@pku.edu.cn; songzd@pku.edu.cn; Liuyang02@pku.edu.cn

**Device fabrication.**
Our devices consist of three layers of helical twisted Bernal bilayer graphene sandwiched by top and bottom hBN and graphite gate. Graphene and hBN (5-30nm thick) are exfoliated on $O_2$ plasma cleaned $SiO_2$/Si substrate. CTTBG stacks were fabricated by 'cut & stack' standard dry transfer technique. Bilayer graphene flake was lithographed into different pieces using a femtosecond laser (HR-Femto-Sci, with a central wavelength ~517nm, pulse width ~ 150fs, repetition frequency ~ 80M and a maximum power ~150mW) focused by a 100X microscope objective. A PC (poly bisphenol A carbonate)/ PDMS (polydimethylsiloxane) stamp was used to pick up top graphite, top hBN and three pieces of bilayer graphene. The whole stack was then released onto a hBN/bottom graphite stack at a PC melting temperature of ~170°C$SiO_2$/Si substrates. The hBN/bottom graphite component was prepared using a PPC (poly propylene carbonate)/PDMS stamp and annealed in Ar/$H_2$ atmosphere at 350°C to remove chemical residues. The samples were then patterned into Hall bar geometry by electron beam lithography and ion etched in $CHF_3$/$O_2$ atmosphere. Electrical edge contacts were then made by chromium(5nm)/gold(70nm).

**Electrical transport measurement.**
Low temperature transport measurement was carried out in a dilution refrigerator with a base temperature about 30mK for device D1 and a cryostat with a base temperature about 1.5K for device D2. Standard low-frequency lock-in techniques were used to measure both devices with gate voltage applied by Keithley 2400 source meter and an excitation of 10 nA at a frequency of 17.777 Hz achieved by SR830 lock-in amplifiers.

**Twist angle determination.**
According to the band structure calculations, a large moiré band gap is expected at $v=\pm 4$ and D = 0V/nm. Thus, we use the large resistance peak at $v=\pm 4$ to determine twist angle for each device. The relation between carrier density $n$ at $v=4$ and twist angle $\theta$ is $n_{v=\pm 4} \sim \pm 8\theta^2/\sqrt{3}a^2$ at small angle limit, where $a = 0.246$ nm is the graphene lattice constant. For device D1, the carrier density for $v = \pm 4$ is $n_{v=\pm 4} \sim \pm 7.09\times 10^{12}$ cm$^{-2}$, which corresponds to $\theta \sim 1.75°$. For device D2, the carrier density for $v = \pm 4$ is $n_{v=\pm 4} \sim \pm 4.94\times 10^{12}$ cm$^{-2}$, which corresponds to $\theta \sim 1.46°$. The applied DC voltage on top gate $V_{tg}$ and bottom gate $V_{bg}$ enable us to independently control carrier density $n = (C_{tg}V_{tg}+C_{bg}V_{bg})/e$ and displacement field $D = (C_{tg}V_{tg}-C_{bg}V_{bg})/2\varepsilon_0$, where $C_{tg}$ ($C_{bg}$) is the capacitance of top (bottom) gate per unit area and $\varepsilon_0$ is the vacuum permittivity. For device D1, $C_{tg}$ and $C_{bg}$ are determined by the voltage difference of quantum Hall states (Fig. S2). For device D2, $c_{tg}$ and $C_{bg}$ are determined by fitting the $R_{xx}$ minimum of the Landau fan diagram, where the measured area contains only single Bernal bilayer graphene (Fig. S4).

**Hartree-Fock calculation.**
We use the continuum model for CTTBG and perform a projected Hartree-Fock calculation on the flat bands. The non-interacting Hamiltonian in each unrotated bilayer with nearest neighbor coupling is

$$h_{\alpha\beta}^{(\eta,i)}(\mathbf{k}) = \begin{pmatrix} v_F\left(\mathbf{k}-\mathbf{K}_i^\eta\right)\cdot\boldsymbol{\sigma}^\eta & 0 \\ 0 & v_F\left(\mathbf{k}-\mathbf{K}_i^\eta\right)\cdot\boldsymbol{\sigma}^\eta \end{pmatrix}_{\alpha\beta} + \begin{pmatrix} \Delta & 0 & 0 & \gamma_1 \\ 0 & 0 & 0 & 0 \\ 0 & 0 & 0 & 0 \\ \gamma_1 & 0 & 0 & \Delta \end{pmatrix}_{\alpha\beta}$$

where we have chosen the Bloch bases of $A_1, B_1, A_2, B_2$, which labels the A,B sublattices on upper and lower layers, $\eta = \pm$ labels the valley, $\mathbf{K}_i^\eta = \eta\frac{4\pi}{3a_0}(1,0), a_0 = 2.46\text{Å}$, and $\boldsymbol{\sigma}^\eta = (\eta\sigma_x,\sigma_y)$, $\sigma_{x,y}$ is the Pauli matrix. We use the parameters $\gamma_1 = 361$meV, $\Delta = 15$meV, $v_F = 5560$ meV·Å from [1].
Following Bistritzer&MacDonald [2], we add the moiré coupling between different bilayers

$$\widehat{H}_{BM} = \sum_{\eta,i} h_{\alpha\beta}^{\eta,i}(R_{-\theta_i}\mathbf{k}) c^\dagger_{\mathbf{k}i\alpha\eta s}c_{\mathbf{k}i\alpha\eta s} + \sum_\eta \left(\widehat{T}_{12}^{(\eta)} + \widehat{T}_{23}^{(\eta)} + h.c.\right)$$

$$\widehat{T}_{ij}^{(\eta)} = \sum_{\alpha\beta s}\sum_{n=1}^{3}[T_{ij}^n]_{\alpha\beta}e^{i\mathbf{q}_{ij}^{n,\eta}\cdot(\mathbf{r}-\mathbf{d}_{ij})}c^\dagger_{\mathbf{r}i\alpha\eta s}c_{\mathbf{r}j\beta\eta s}$$

$$T_{ij}^n = \begin{pmatrix} 0 & 1 \\ 0 & 0 \end{pmatrix} \otimes \left(w_0\sigma_0 + w_1\left[\cos\frac{2\pi(n-1)}{3}\sigma_x + \sin\frac{2\pi(n-1)}{3}\sigma_y\right]\right)$$

with $w_0 = 110$meV, $w_1 = 88$meV, where $R_\theta$ denotes rotation $\theta$ anti-counterwisely, $\theta_i = (\theta, 0, -\theta)$, $\mathbf{q}_{12}^{n,\eta} = \mathbf{q}_{23}^{n,\eta} = R_{(n-1)\frac{2\pi}{3}}\mathbf{q}^\eta$, $\mathbf{q}^\eta = \eta\frac{8\pi\sin\frac{\theta}{2}}{3\sqrt{3}}(0,-1)$, $\mathbf{d}_{ij}$ controls the displacement of the moiré patterns.

Besides, the displacement field Hamiltonian $\hat{H}_d$ brings a potential $V_D\left(l-\frac{7}{2}\right)$ to the $l$ th monolayer graphene ($l=1,\ldots 6$) where we use $V_D = 5\text{meV}$. We diagonalize the non-interacting Hamiltonian $\hat{H}_0 = \hat{H}_{BM} + \hat{H}_d$ to obtain the band structure in Fig.1 d and the wavefunctions of flat bands. We denote the creation operators for the two flat bands as $c^\dagger_{\mathbf{k}m\eta s}$, $m=1,2$ and their energies as $\epsilon_{\mathbf{k}m\eta}$.

We then add the interaction Hamiltonian and project the total Hamiltonian to the two flat bands following [3], which yields

$$\hat{H} = \sum_{m\eta s}\sum_{\mathbf{k}\in MBZ} \epsilon_{\mathbf{k}m\eta} c^\dagger_{\mathbf{k}m\eta s} c_{\mathbf{k}m\eta s}$$

$$+ \frac{1}{2} \sum_{\mathbf{k}\mathbf{k}'\mathbf{q}\in MBZ} \sum_{\substack{mm'nn' \\ \eta\eta' ss'}} U^{\eta\eta'}_{mn,m'n'}(\mathbf{q};\mathbf{k},\mathbf{k}') : c^\dagger_{\mathbf{k}+\mathbf{q}m\eta s} c_{\mathbf{k}n\eta s} :: c^\dagger_{\mathbf{k}'-\mathbf{q}m'\eta' s'} c_{\mathbf{k}'n'\eta' s'} :$$

$U^{\eta\eta'}_{mn,m'n'}(\mathbf{q};\mathbf{k},\mathbf{k}')$ is the matrix element of the interaction on the flat bands, $:c^\dagger_{\mathbf{k}+\mathbf{q}m\eta s} c_{\mathbf{k}n\eta s}: = c^\dagger_{\mathbf{k}+\mathbf{q}m\eta s} c_{\mathbf{k}n\eta s} - \frac{1}{2}\delta_{\mathbf{q}0}\delta_{mn}$. In this article, we adopt the double-gate-screened interaction $V(\mathbf{q}) = \frac{e^2}{2\epsilon|\mathbf{q}|}\tanh(\xi|\mathbf{q}|/2)$ where $\xi = 30\text{nm}$ is the distance between the two gates and $\epsilon = 6$ is the relative dielectric constant of hBN.

Besides, the displacement field Hamiltonian $\hat{H}_d$ brings a potential $V_D\left(l-\frac{7}{2}\right)$ to the $l$ th monolayer graphene ($l=1,\ldots 6$) where we use $V_D = 5\text{meV}$. This is an effective interlayer potential containing both the external displacement field from the gate and the internal charge redistribution, and we neglect the possible non-uniform potential difference between layers for simplicity, as done in [4].

To study the possible translation symmetry breaking order, we enlarge the unit cells, i.e. consider the folded moiré Brillouin zones(fMBZ) with a set of vectors $\mathbf{Q}_b$, by which we can rewrite the momentum in the moiré Brillouin zones as $\mathbf{k} + \mathbf{Q}_b$ where $\mathbf{k} \in$ fMBZ. The Hartree-Fock order parameters is then $O_{bm\eta s;b'm'\eta's'}(\mathbf{k}) = \left\langle c^\dagger_{\mathbf{k}+\mathbf{Q}_b m\eta s} c_{\mathbf{k}+\mathbf{Q}_{b'} m'\eta' s'}\right\rangle - \frac{1}{2}\delta_{bb'}\delta_{mm'}\delta_{\eta\eta'}\delta_{ss'}$. Within the standard Hartree-Fock decomposition, the interaction Hamiltonian is approximated by

$$\hat{H}^{MF}_I = \frac{1}{\Omega_{tot}} \sum_{\mathbf{k}\in fMBZ} \sum_{bb'} \sum_{m\eta s, m'\eta' s'} \Delta_{bm\eta s, b'm'\eta' s'}(\mathbf{k}) c^\dagger_{\mathbf{k}+\mathbf{Q}_b m\eta s} c_{\mathbf{k}+\mathbf{Q}_{b'} m'\eta' s'}$$

where

$$\Delta_{bm\eta s, b'm'\eta' s'}(\mathbf{k}) =$$

$$\sum_{b''}\sum_{nn'}\left(\delta_{ss'}\delta_{\eta\eta'}\sum_{\eta'' s''} U^{(\eta\eta'')}_{mm',nn'}(\mathbf{Q}_b - \mathbf{Q}_{b'}; \mathbf{k}+\mathbf{Q}_{b'}, \mathbf{k}'+\mathbf{Q}_{b''}) O_{b''-b+b'n\eta'' s''; b''n'\eta'' s''}(\mathbf{k}')\right.$$

$$\left. - \sum_{\eta''} U^{(\eta\eta'')}_{mn',nm'}\left(\begin{array}{c}\mathbf{k}+\mathbf{Q}_b - \mathbf{k}' - \mathbf{Q}_{b''}; \mathbf{k}'+\mathbf{Q}_{b''}, \mathbf{k} \\ +\mathbf{Q}_{b'}\end{array}\right) O_{b''-b+b',n\eta' s'; b''n'\eta s}(\mathbf{k}')\right)$$

and we have defined $b \pm b'$ by $\mathbf{Q}_{b\pm b'} \equiv \mathbf{Q}_b \pm \mathbf{Q}_{b'}$ mod reciprocal lattice vector of MBZ.

We start from randomized initial value of order parameters and perform self-consistent calculations to obtain convergent order parameters and energy, and the Hartree-Fock bands in

Fig.3 d. The calculated band gap is larger than the experimental extracted value, which is not peculiar since the Hartree-Fock method always overestimates the band gap.

The calculated non-interacting band structures are displayed in Fig. S1. For both twist angles we have discussed in main text, well isolated flat bands have been captured. When applying perpendicular displacement field, a band gap is induced at charge neutral point which have is consistent with our experimental results. Significantly, the bandwidth remains almost the same when changing twist angle near $\theta=1.7°$. The similarity of the band structures indicates that the picture of CCDWI states in both devices is the same. Fig. S1d shows the bandwidth versus twist angle for ABABBC stacking order which is one of the two possible stacking orders (the other one- ABABAB is adopted in our calculation). The bandwidth of ABABBC stacked CTTBG can be ~40meV which is in sharp contrast with ABABAB stacking order and also conflicting with our experiment.

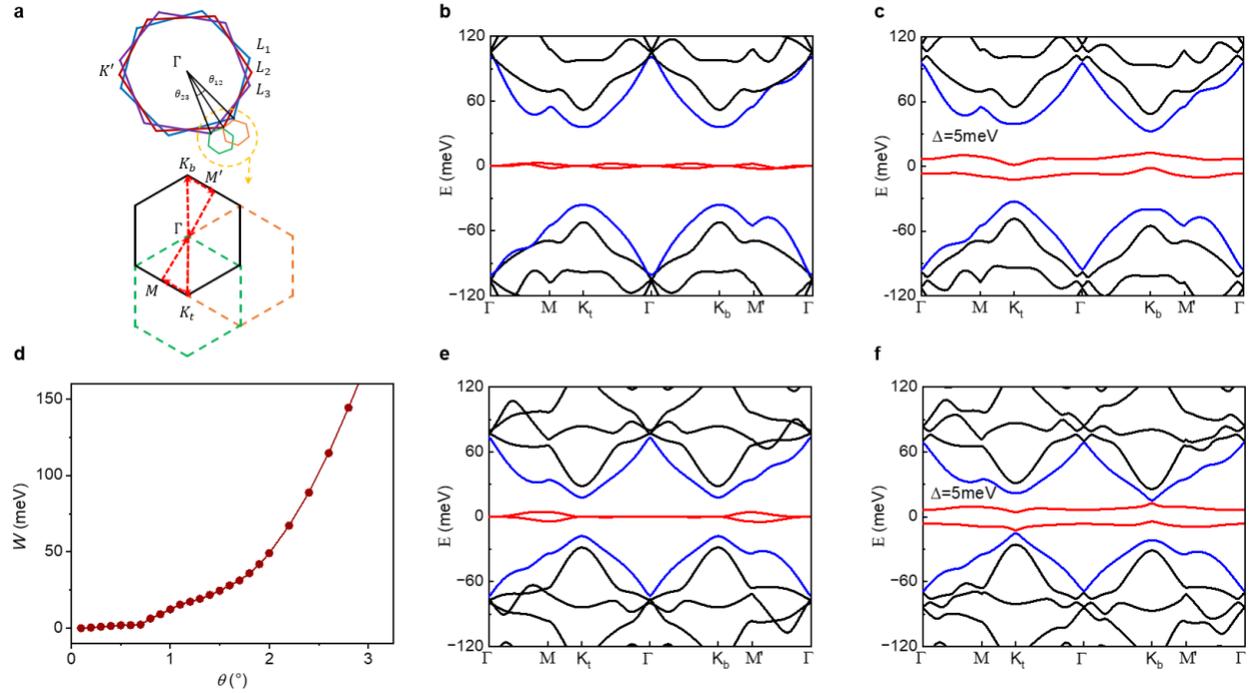

**Fig.S1|Band structures of CTTBG. a,** Schematic of the Brillouin zone of CTTBG, the dashed arrows denote the projection path. The black hexagon represents for the Brillouin zone of the non-interacting calculation. **b-c,** Band structure of CTTBG at $\theta=1.7°$ with **(c)** and without **(b)** perpendicular displacement field achieved by non-interacting calculation for ABABAB stacking order. Δ=5meV represents for the electrical potential energy between top and bottom hBN. **d,** The bandwidth of the central flat band as a function of twist angle $\theta$ for ABABBC stacking order. **e-f,** Band structure of CTTBG at $\theta=1.46°$ with **(f)** and without **(e)** perpendicular displacement field achieved by non-interacting calculation.

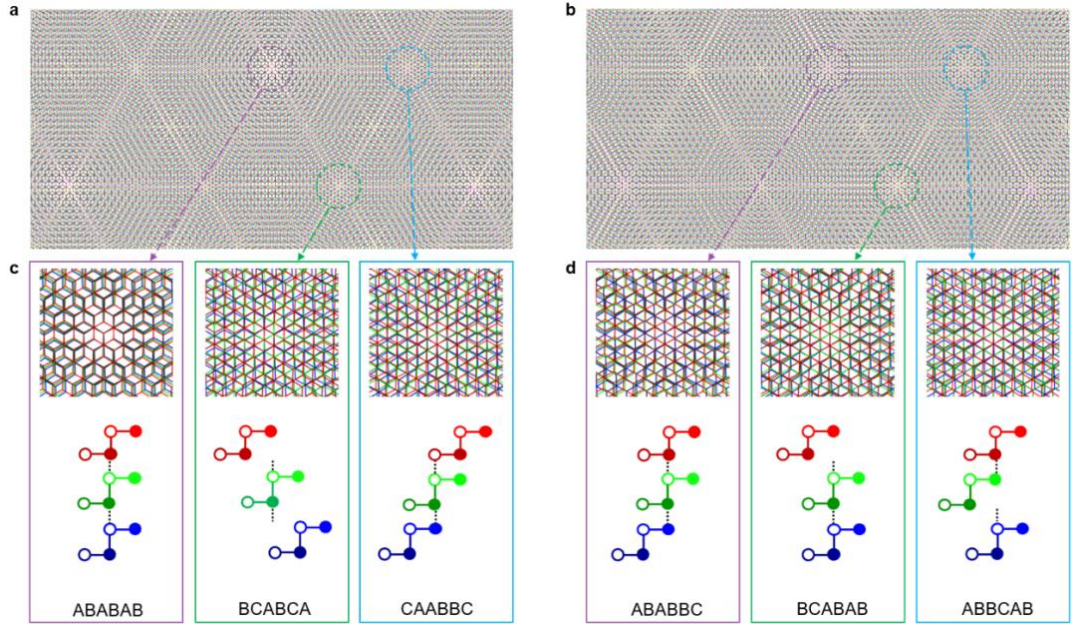

**Fig.S2|Schematic of ABABAB and ABABBC stacking order. a-b,** Moiré structure of CTTBG for ABABAB **(a)** stacking order and ABABBC **(b)** stacking order. The dashed circles indicate different stacking sites. **c-d,** Zoom-in moiré structure of different stacking sites for ABABAB moiré stacking order **(c)** and ABABBC stacking order **(d)**. The bottom schematics show the side-looking of the moiré structure.

Fig. S3 displays more details of our devices. For device D1, an ac excitation of 10nA is applied through contacts JF with $V_{bg}$= 6V. Three pairs of contacts show nearly identical behavior. The gate voltages for each correlated insulating state or band insulating state are almost the same for each pair of contact, indicating very high homogeneity in our device. For this work, we focus on contacts DE to measure longitudinal resistance and DG to measure Hall resistance. For device D2, 10nA ac current is applied through contacts AG and $V_{bg}$ is set to be -4V. The behavior of correlated insulating states and band insulating states also show great consistency for contacts BC and CD, which indicates device D2 is homogeneous in corresponding device area. Thus, we focus on contacts CD to measure longitudinal resistance and contacts CL to measure Hall resistance.

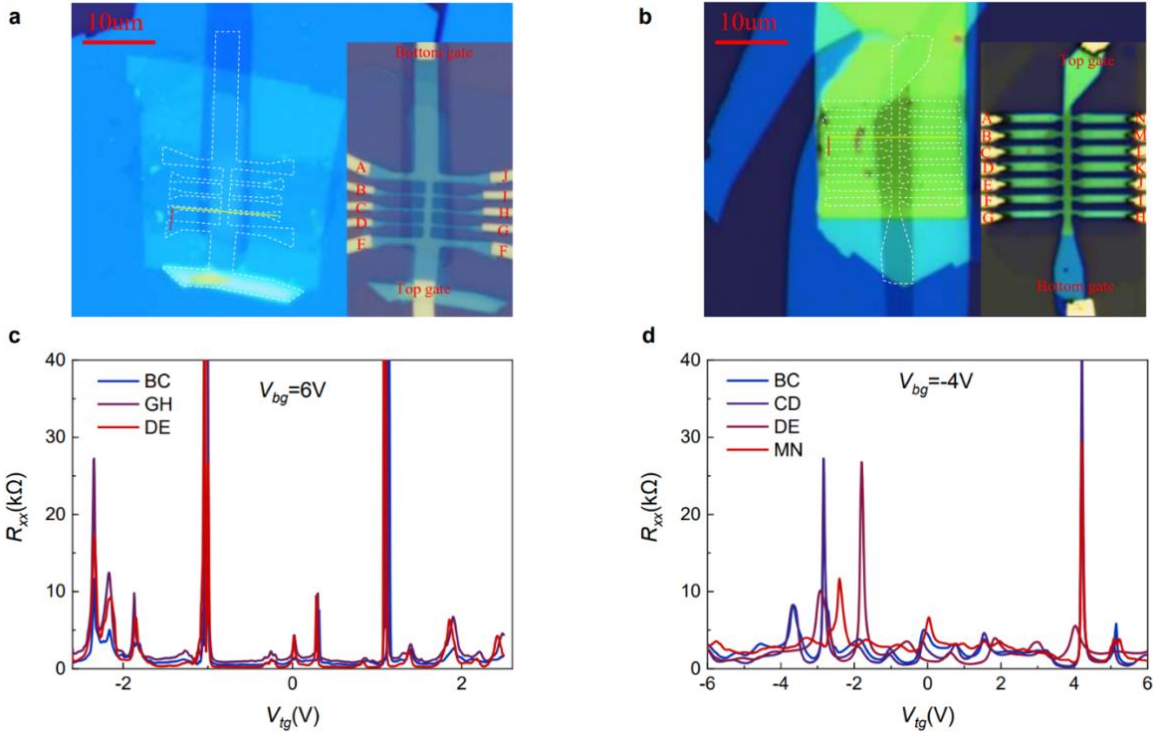

**Fig. S3|CTTBG devices. a-b,** Optical images of CTTBG stacks of device D1**(a)** and device D2 **(b)** after. The red and yellows lines represent for the contact pair we focus on $R_{xx}$ and $R_{xx}$ respectively. The inlet shows CTTBG devices after nanofabrication procedures. **c-d,** Longitudinal resistance as a function of $V_{tg}$ for different contact pairs.

Fig. S4 shows the measurement of quantum Hall effect of device D1 which is used to obtain the gate efficiency. $R_{xy}$ show accurate quantization while $R_{xx}$ exhibit significant minimum at the same gate voltage when $R_{xx}$ show quantization for Chern number $C$=1, 2. Based on quantum Hall effect $\Delta C = h\Delta n_{2D}/eB_\perp = hC_{tg}\Delta V_{tg}/eB_\perp$, the top gate capacitance $C_{tg}$ is calculated by the average difference between $C$ =1, 2 states defined by resistance dips. The ratio of top gate capacitance and bottom gate capacitance is extracted from the position of charge neutral point in the dual-gate map in Fig.1e ($C_{tg}$ =532nF/cm$^2$, $C_{bg}$ =95nF/cm$^2$).

Fig.S5 shows the $R_{xx}$ as a function of $V_{tg}$ and $B_\perp$ measured from contact IH which in device D2 and a Landau Fan diagram is clearly observed. The red dashed lines indicate the position of Landau level gaps. Considering spin-valley degeneracy of graphene, the observed Landau level should host four-fold degeneracy. Thus, the difference of Chern number in adjacent Landau levels are four $\Delta C$=4. The value of $C_{tg}$ can be obtained from $C_{tg}\Delta V_{tg}=\Delta C*B_\perp e/h$. With similar method used in device D1, the value of $C_{tg}$ and $C_{bg}$ are deduced to be $C_{tg}$=113nF/cm$^2$, $C_{bg}$=116nF/cm$^2$. The thickness of bottom hBN has been measured by atom force microscopy d$_{bottom}$=24.2nm. By $C_{bg}=\varepsilon\varepsilon_0/$d$_{bottom}$, where $\varepsilon_0$ is the vacuum permittivity and $\varepsilon$ is the relative permittivity of hBN, the value of $C_{bg}$ is deduced to be 117 nF/cm$^2$ ($\varepsilon$=3.2), which is consistent with the result obtained from Landau levels.

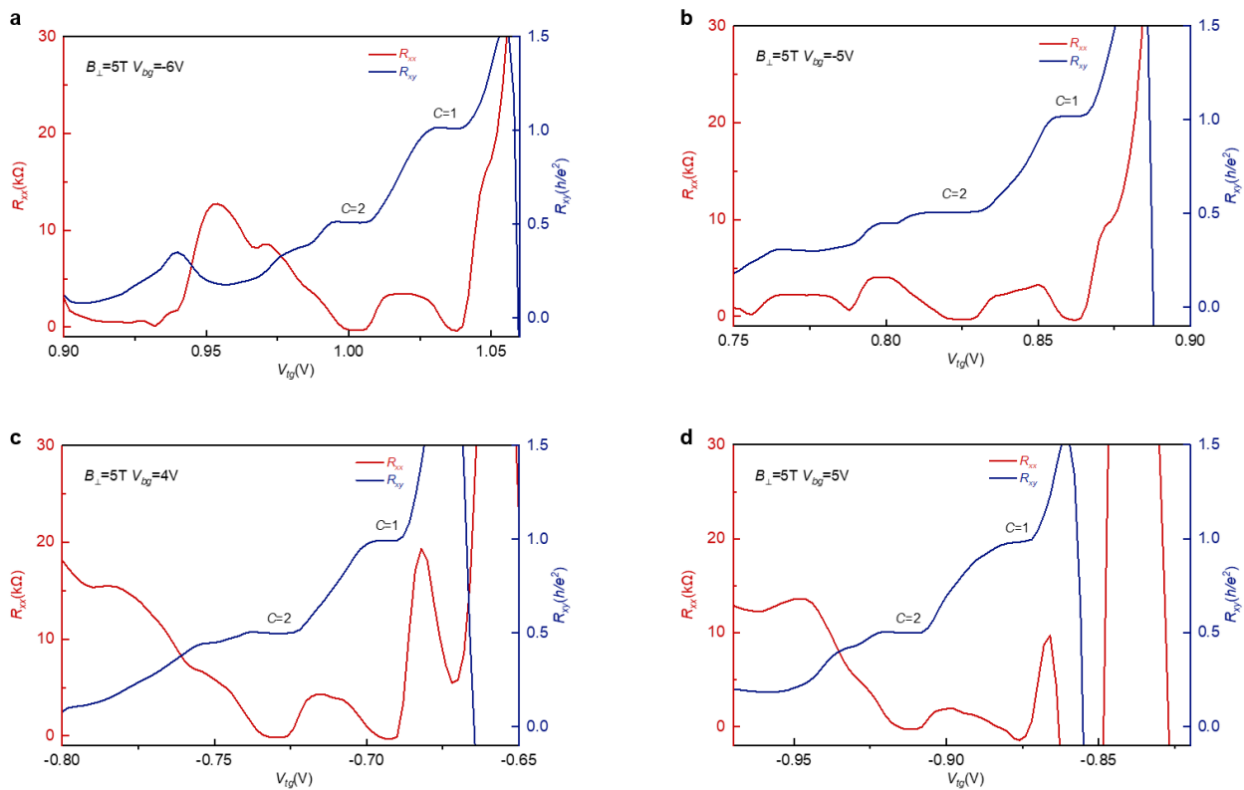

**Fig. S4|Quantum Hall effect of device D1.** Longitudinal resistance $R_{xx}$ and Hall resistance $R_{xy}$ as a function of top gate voltage $V_{tg}$ at perpendicular magnetic field $B_\perp$ and different bottom gate voltage $V_{bg}$.

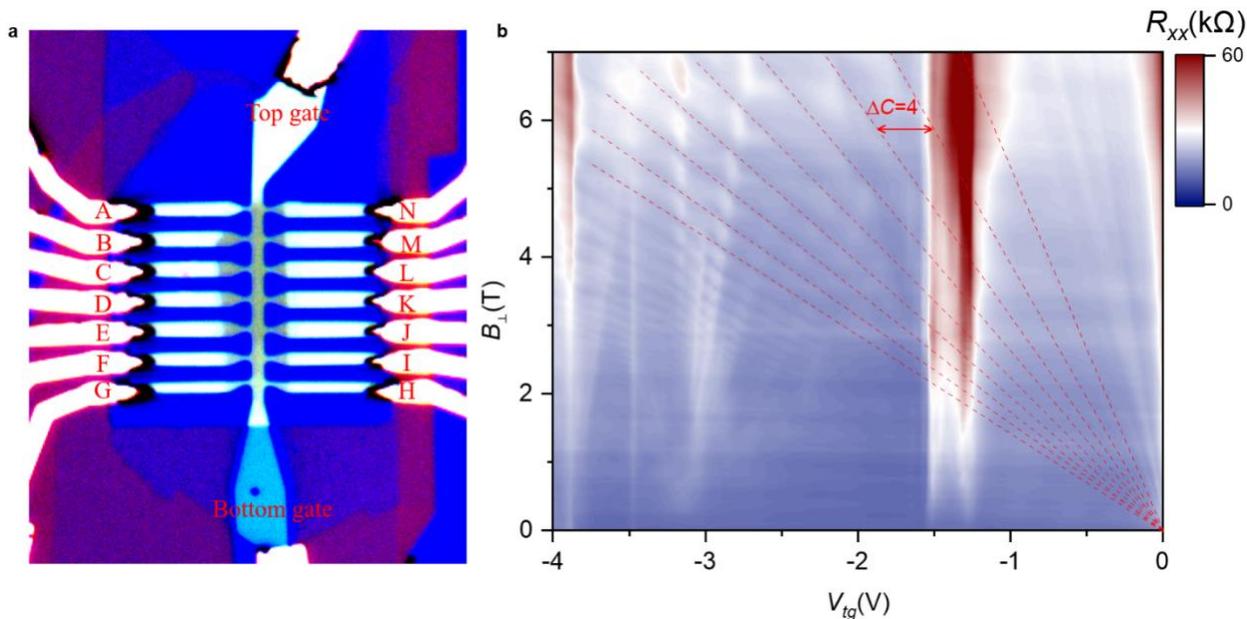

**Fig.S5|Landau Fan diagram of device D2. a,** Optical image to illustrate the measured device area. The upper half of the measured area contains the first and second transferred bilayer graphene. The bottom half of the measured area only contains the first transferred bilayer graphene, the upper

half of the measured area only contains the first and second transferred bilayer graphene. **b,** The dependence of longitudinal resistance on top gate voltage $V_{tg}$ and perpendicular magnetic field $B_\perp$, 10nA ac current is applied through contacts NH and the voltage of contacts IH is measured. Red dashed lines denote the resistance dips which correspond to Landau levels gaps.

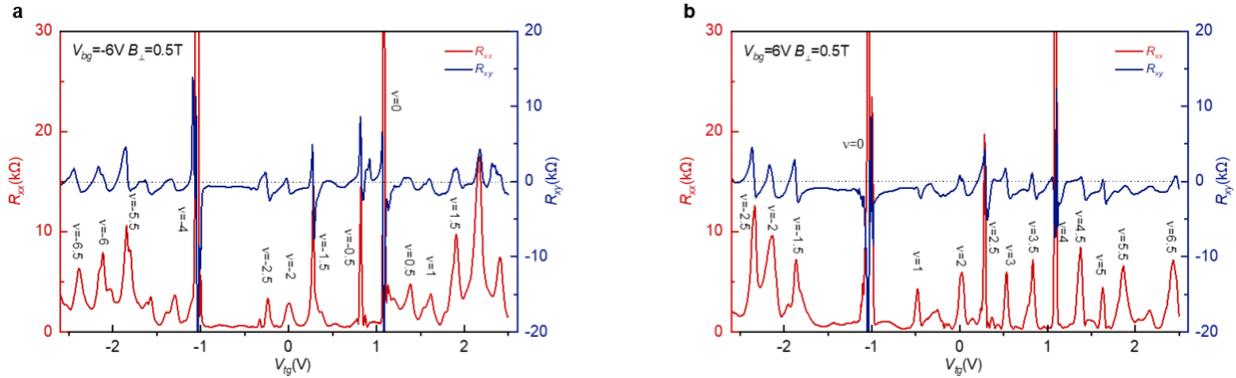

**Fig.S6|Comparison for $R_{xx}$ and $R_{xy}$.** Longitudinal resistance $R_{xx}$ and Hall resistance $R_{xy}$ as a function of top gate voltage $V_{tg}$ at bottom gate voltage $V_{bg}$=-6V **(a)** and $V_{bg}$=6V for device D1 **(b)**. $R_{xy}$ exhibits significant sign change at both integer filled and half integer filled states. The sign change of $R_{xy}$ provides another evidence for the existence of correlated gaps at both integer filled and half integer filled states apart from thermal activation behavior.

Fig. S7-8 shows the details of thermal activation behavior of all gaps we discussed in main text for device D1 and D2 respectively. All the marked filled states show significant thermal behavior. All of the $v=\pm 4$ and $v=0$ filled states show profound large resistance peak, indicating large band gap at $v=\pm 4$ and band gap opening by displacement field at $v=0$. Most of $v=\pm 4$ and $v=0$ filled states show weak dependence on temperature, for the band gaps of these states are too large to exhibit thermal activation at the relatively low temperature.

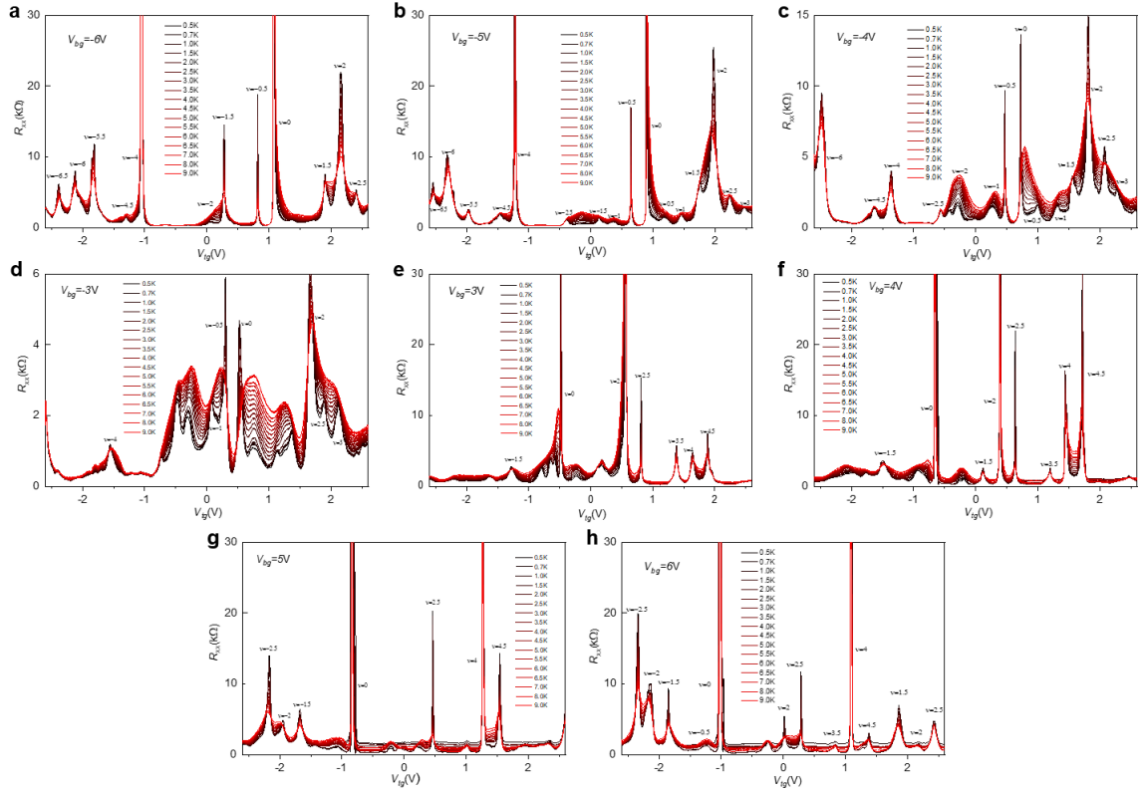

**Fig.S7|$R_{xx}$ at different temperature for device D1. a-h,** Longitudinal $R_{xx}$ resistance as a function of top gate voltage $V_{tg}$ at different temperature and bottom gate voltage $V_{bg}$=-6V **(a)**, $V_{bg}$=-5V **(b)**, $V_{bg}$=-4V **(c)**, $V_{bg}$=-3V **(d)**, $V_{bg}$=3V **(e)**, $V_{bg}$=4V **(f)**, $V_{bg}$=5V **(g)**, $V_{bg}$=6V **(h)** for device D1.

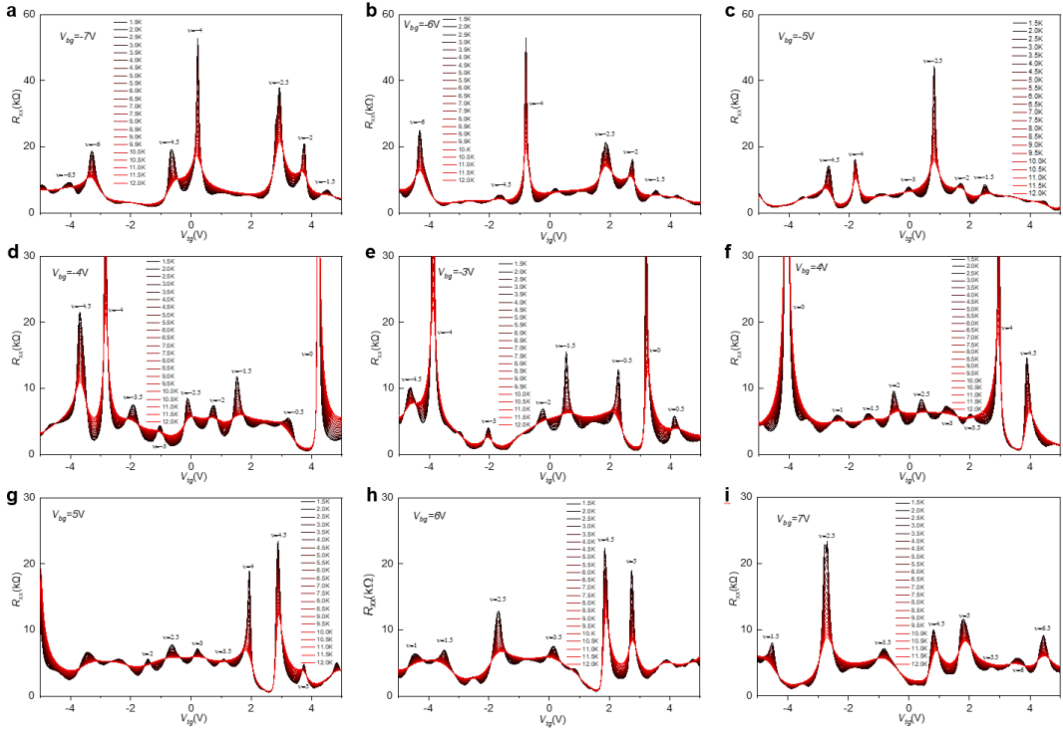

**Fig.S8|$R_{xx}$ at different temperature for device D2. a-i,** Longitudinal $R_{xx}$ resistance as a function of top gate voltage $V_{tg}$ at different temperature and bottom gate voltage $V_{bg}$=-7V **(a)**, $V_{bg}$=-6V **(b)**, $V_{bg}$=-5V **(c)**, $V_{bg}$=-4V **(d)**, $V_{bg}$=3V **(e)**, $V_{bg}$=4V **(f)**, $V_{bg}$=5V **(g)**, $V_{bg}$=6V **(h)**, $V_{bg}$=7V **(h)** for device D2.

For $v$=2, the quantity of activation gap is $\Delta$~3.85meV. For $v$=-1/2, the quantity of activation gap is $\Delta$~1.51meV (D=0.568V/nm), $\Delta$~1.83meV (D=0.462V/nm), $\Delta$~0.98meV (D=0.355V/nm). For $v$=5/2, the quantity of activation gap is $\Delta$~0.72meV (D=-0.236V/nm), $\Delta$~1.01meV (D=-0.129V/nm), $\Delta$~0.80meV (D=0.023V/nm), $\Delta$~0.81meV (D=0.083V/nm), $\Delta$~1.45meV (D=0.189V/nm). ln($R_{xx}$) shows apparent linear dependence on $1/T$ at fitted area, which means the fitting of the thermal activation is reliable. Based on the activation behavior, it is apparent that the gap of half-filled state $v$=2 is much larger than the gap of half-integer filled states.

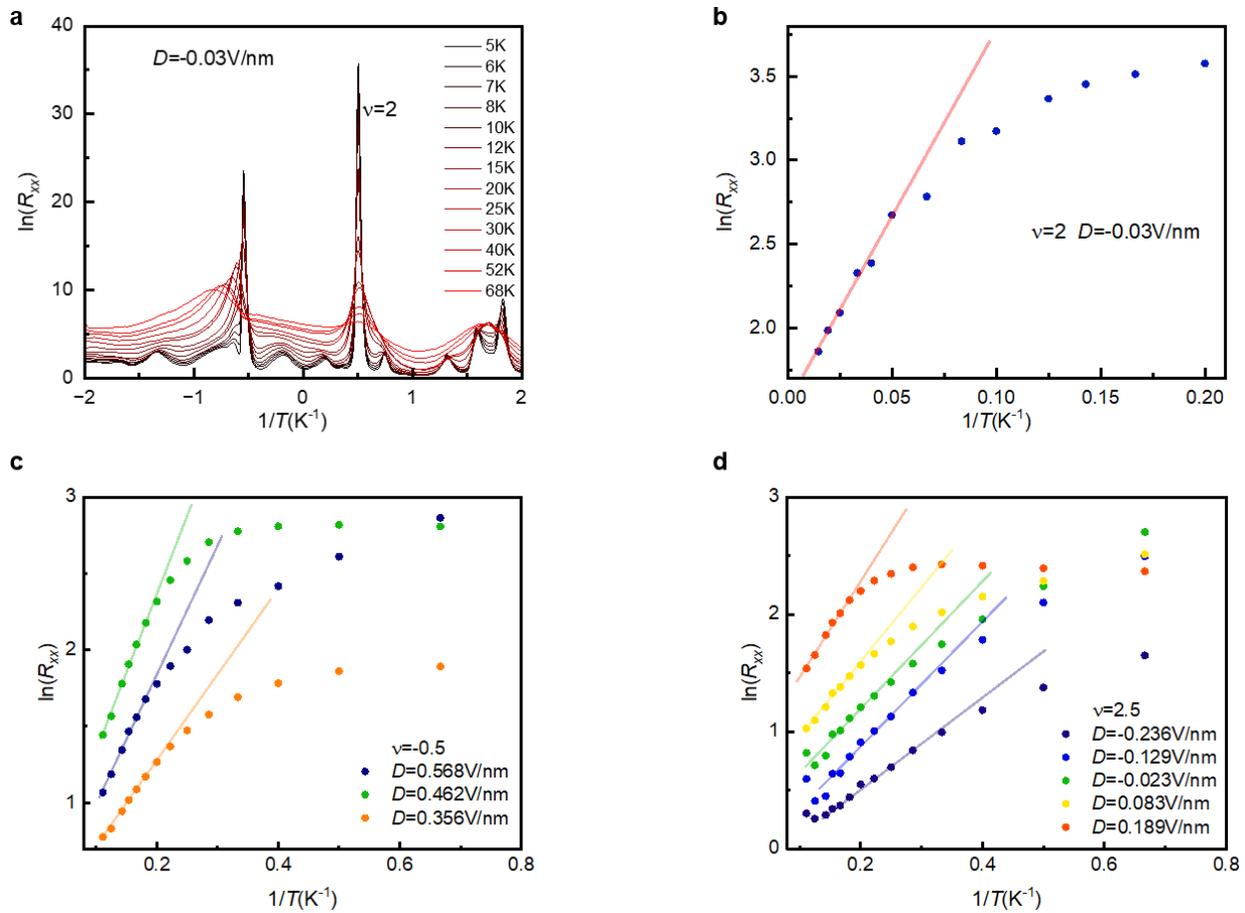

**Fig.S9|Fitting of activation gaps. a,** Longitudinal resistance $R_{xx}$ as a function of top gate voltage $V_{tg}$ at bottom gate voltage $V_{bg}$=3.5V at temperature varying from 5K to 68K for device D1. The measurement at larger temperature is needed since the activation gap at $v$=2 is larger. **b-d,** The logarithm of longitudinal resistance ln($R_{xx}$) plotted against inverse temperature $1/T$ at $v$=2 **(b)**, $v$ -1/2 **(c)** and $v$=5/2 **(d)** for device1. The straight line is fit to $R_{xx} \propto \exp(\Delta/2kT)$ for temperature activation behavior.

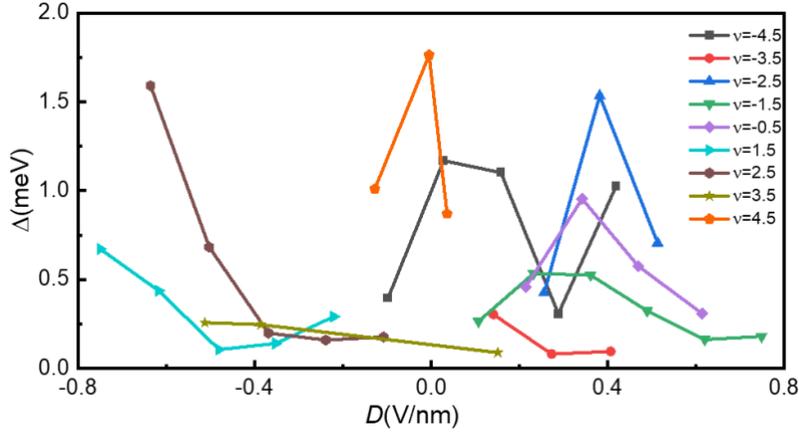

**Fig.S10| Activation gap of half-integer filled states of device D2.** Dependence of correlated half integer filled state gaps on displacement field for device D2. Consistently to device D1, the correlated half integer-filled gaps can be tuned by displacement field and the quantity of the gaps are similar.

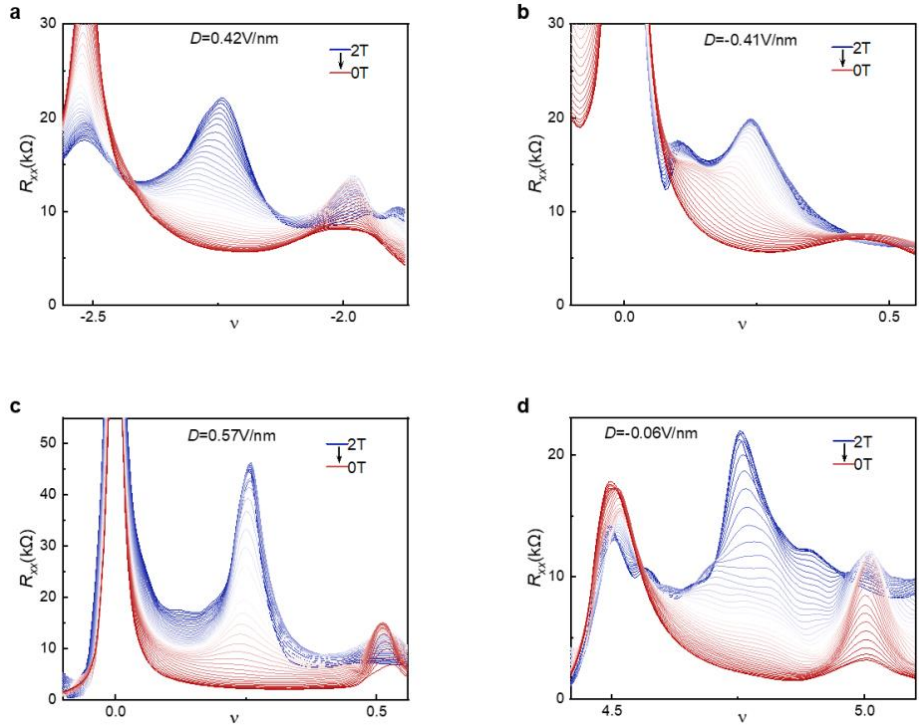

**Fig.S11| Emergence of quarter filled states for device D2.** Dependence of longitudinal resistance $R_{xx}$ on filling factor $v$ and perpendicular magnetic field $B_\perp$ with different displacement field. Similar to the behavior of device1 in Fig.2e-h, as $B_\perp$ is gradually increased, the quarter integer filled states are formed ($v=-9/4$, $v=7/4$, $v=-1/4$, $v=1/4$, $v=19/4$).

Fig.S12-13 show Longitudinal resistance $R_{xx}$ and Hall resistance $R_{xy}$ as a function of filling factor $v$ and displacement field $D$ with different perpendicular magnetic field $B_\perp=0.5T$, 1T, 1.5T, 2T for device D1 and device D2 respectively. At low $B_\perp$ field, some of the original integer filled and half-integer filled states are strengthened since the corresponding $R_{xx}$ is enlarged or exist at larger range of displacement field ($v=-0.5$, $v=1.5$, $v=2$, $v=2.5$, $v=3.5$, $v=4.5$ for device D1), and new integer filled and half integer filled states are generated ($v=-2.5$, $v=3$, $v=5$ for device D1 and $v=-6$, $v=-4$, $v=-5/2$, $v=-2$, $v=-0.5$, $v=2$, $v=9/2$ for device D2). As the $B_\perp$ field is further enhanced, new quarter filled states are generated ($v=5/4$, $v=7/4$, $v=9/4$, $v=11/4$, $v=17/4$, $v=19/4$ for device D1 and $v=-9/4$, $v=-7/4$, $v=-1/4$, $v=1/4$, $v=19/4$ for device D2). Each integer, half- integer and quarter filled states can be identified from $R_{xy}$ dual gate maps.

As shown in Fig.SI14, $R_{xx}$ at quarter filled states ($v=-9/4$, $1/4$, $19/4$, $21/4$ for device D2) also show thermal activation behavior at $B_\perp=2T$, indicating the formation of correlated insulating states modulated by charge density wave and the opening of correlated gaps.

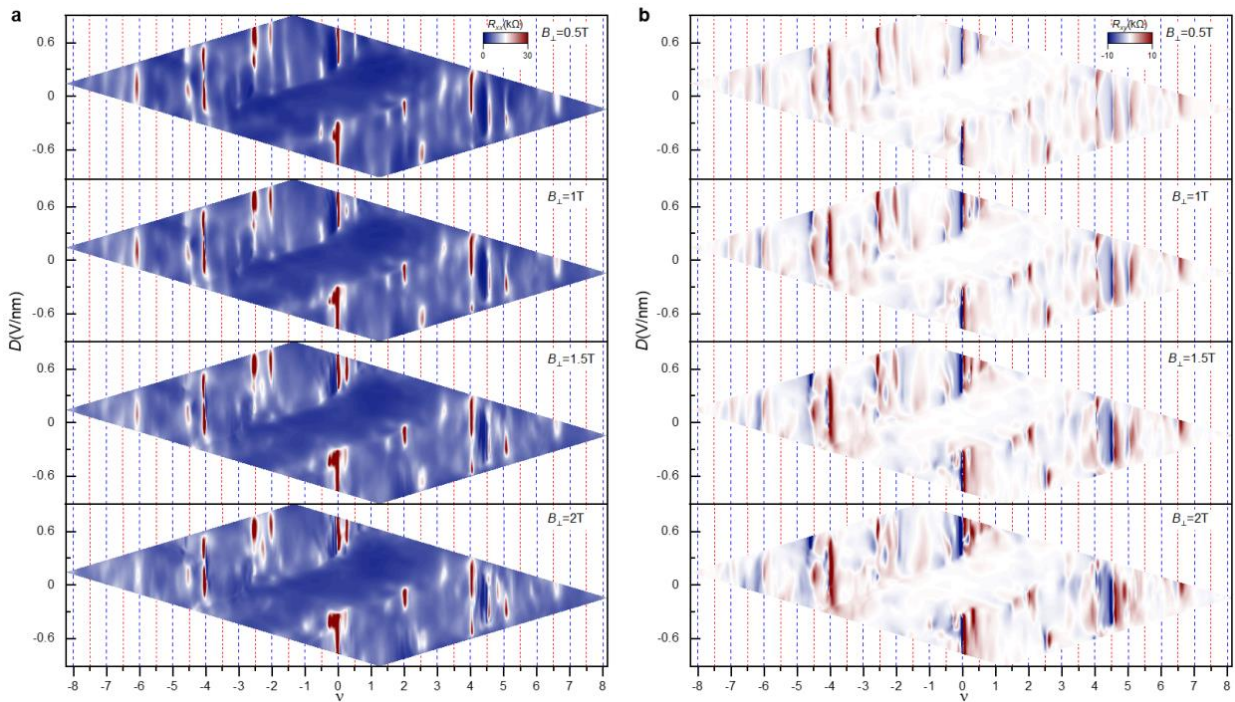

**Fig.S12| Dual-gate 2D maps with $B_\perp$ of device D2.** Longitudinal resistance $R_{xx}$ **(a)** and Hall resistance $R_{xy}$ **(b)** as a function of filling factor $v$ and displacement field $D$ with different perpendicular magnetic field $B_\perp=0.5T$, $1T$, $1.5T$, $2T$.

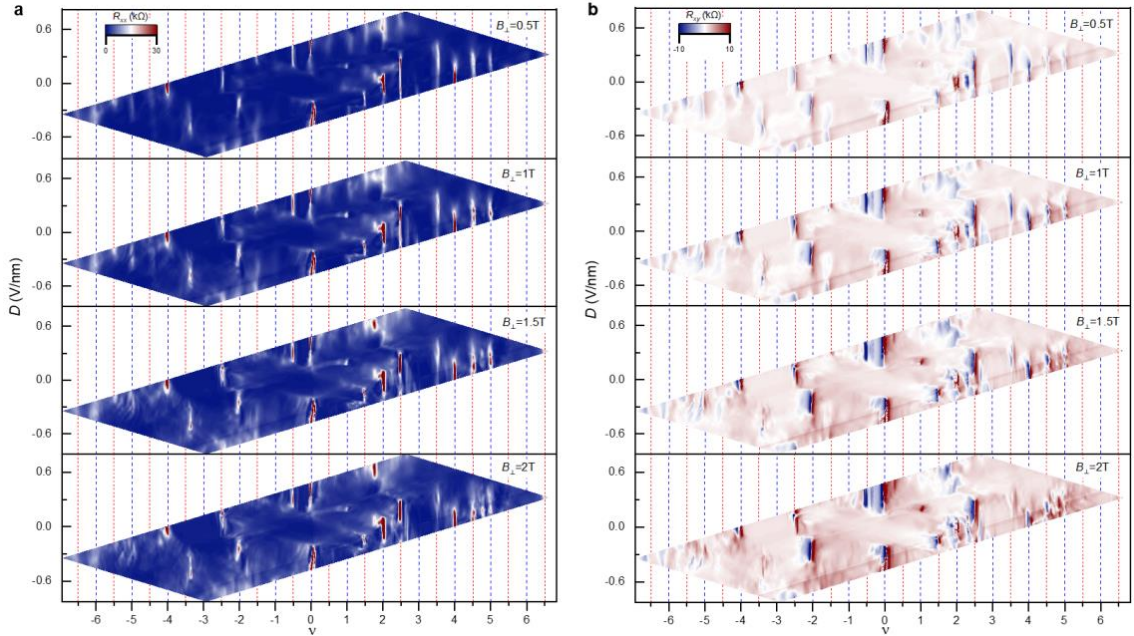

**Fig.S13| Dual-gate 2D maps with $B_\perp$ of device D1.** Longitudinal resistance $R_{xx}$ **(a)** and Hall resistance $R_{xy}$ **(b)** as a function of filling factor $v$ and displacement field $D$ with different perpendicular magnetic field $B_\perp$=0.5T, 1T, 1.5T, 2T.

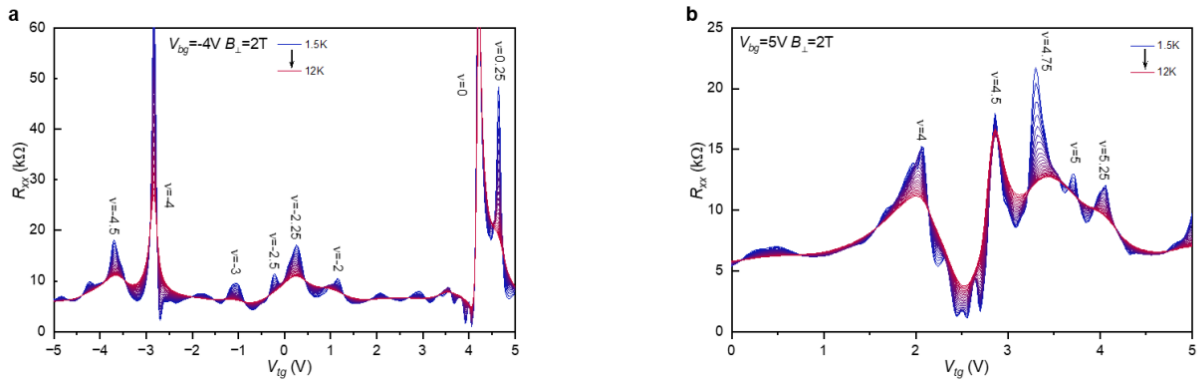

**Fig.S14|Temperature activation behavior at quarter filled states.** Longitudinal resistance $R_{xx}$ as a function of top gate voltage $V_{tg}$ at bottom gate voltage $V_{bg}$=-4V **(a)** and $V_{bg}$=5V **(b)** at temperature varying from 1.5K to 12K and perpendicular magnetic field $B_\perp$=2T for device D2.

Fig.S15 shows the mobility of the CTTBG devices. For the calculation of the mobility, we adopt the equation $\mu=(d\sigma/dn)/e$ (where $\sigma$ is conductivity, $n$ is the carrier density, $e$ is the electric charge of a single electron) and the data at finite displacement field. The calculated mobility for both devices are presented in Fig.S15d-e. The mobility of the used graphene is in the range of ~$2\times10^5$cm$^2$/(V·s).

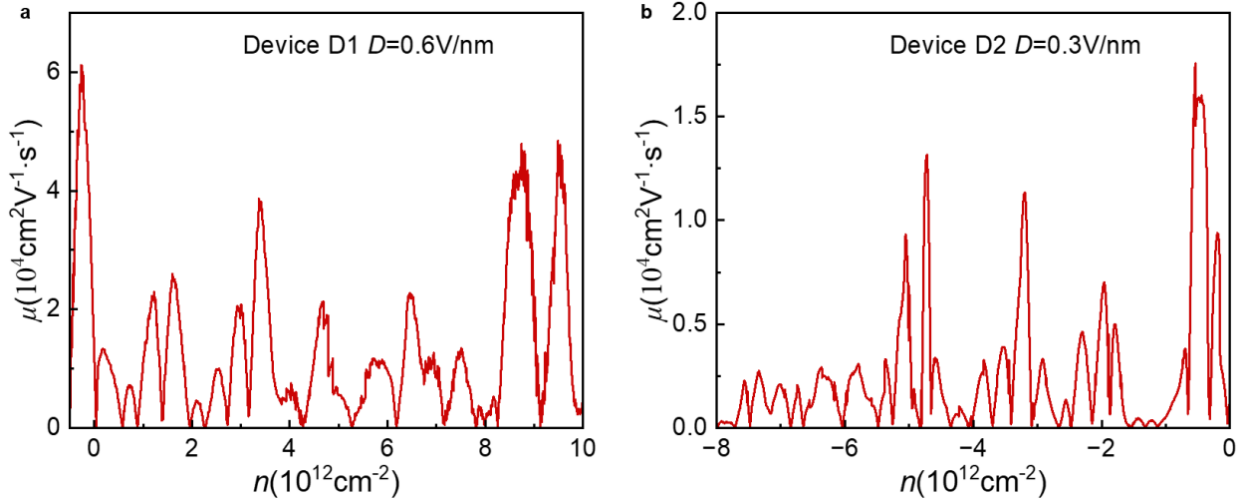

**Fig.S15|Mobility of both devices.** Calculated mobility as a function of carrier density for device D1 at displacement field $D=0.6$V/nm **(a)** and for device D2 at $D=0.3$V/nm **(b)**.

Fig.SI16-17 show the electron distribution by Hartree-Fock calculation in real space for ABABAB stacking order at $v=1.75$ and for ABABBC stacking order at $v=0.5$, $v=1$, $v=1.75$ respectively. For quarter integer filled states, there are two possible ways to form charge density wave. The first way is to form bubble phase, where CDW breaks translation symmetry in two perpendicular directions and multiply the moiré unit cell for two times respectively, totally multiplying the unit cell for four times. The second way is to form strip phase, where CDW breaks translation symmetry in only one direction and multiply the moiré unit cell for four times in this direction. CDW at both the half and quarter integer filled states can also exist at ABABAB stacking order and break translation symmetry in a similar way.

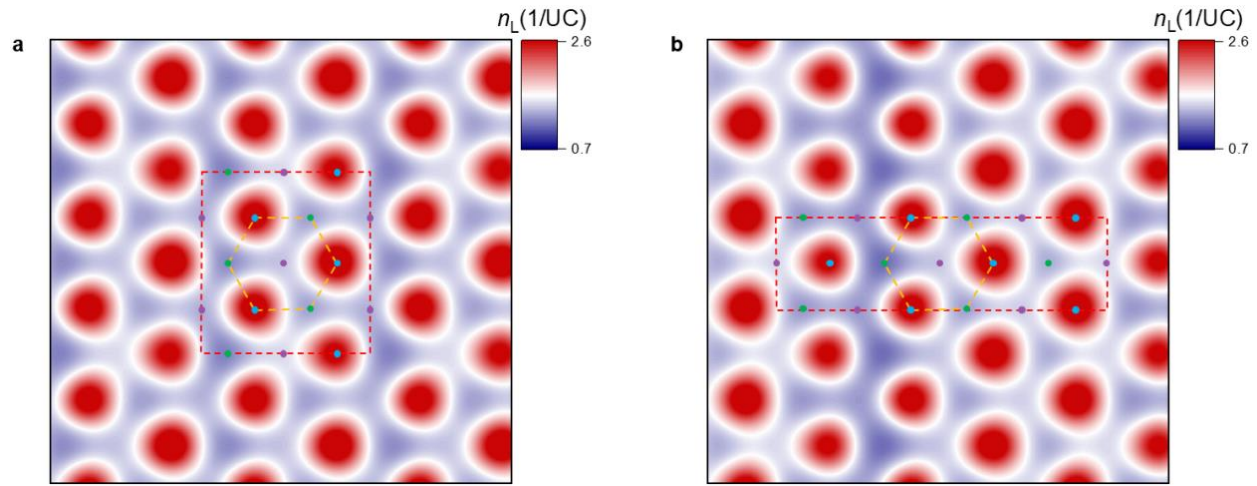

**Fig.S16|Electron distribution at $v=1.75$ for ABABAB stacking order.** Electron distribution by Hartree-Fock calculation in real space at $v=1.75$ for bubble phase **(a)** and strip phase **(b)** for ABABAB stacking order, the value $n_L$ represent for the number of electrons per moiré unit cell.

The circle dots represent for different stacking sites. Purple dots represent for ABABBC stacking sites, blue dots represent for ABBCAB stacking sites, green dots represent for BCABAB stacking sites. The orange hexagonal dashed line represents for the original moiré unit cell, the red rectangular dashed line represents for the expanded moiré unit cell formed by CDW.

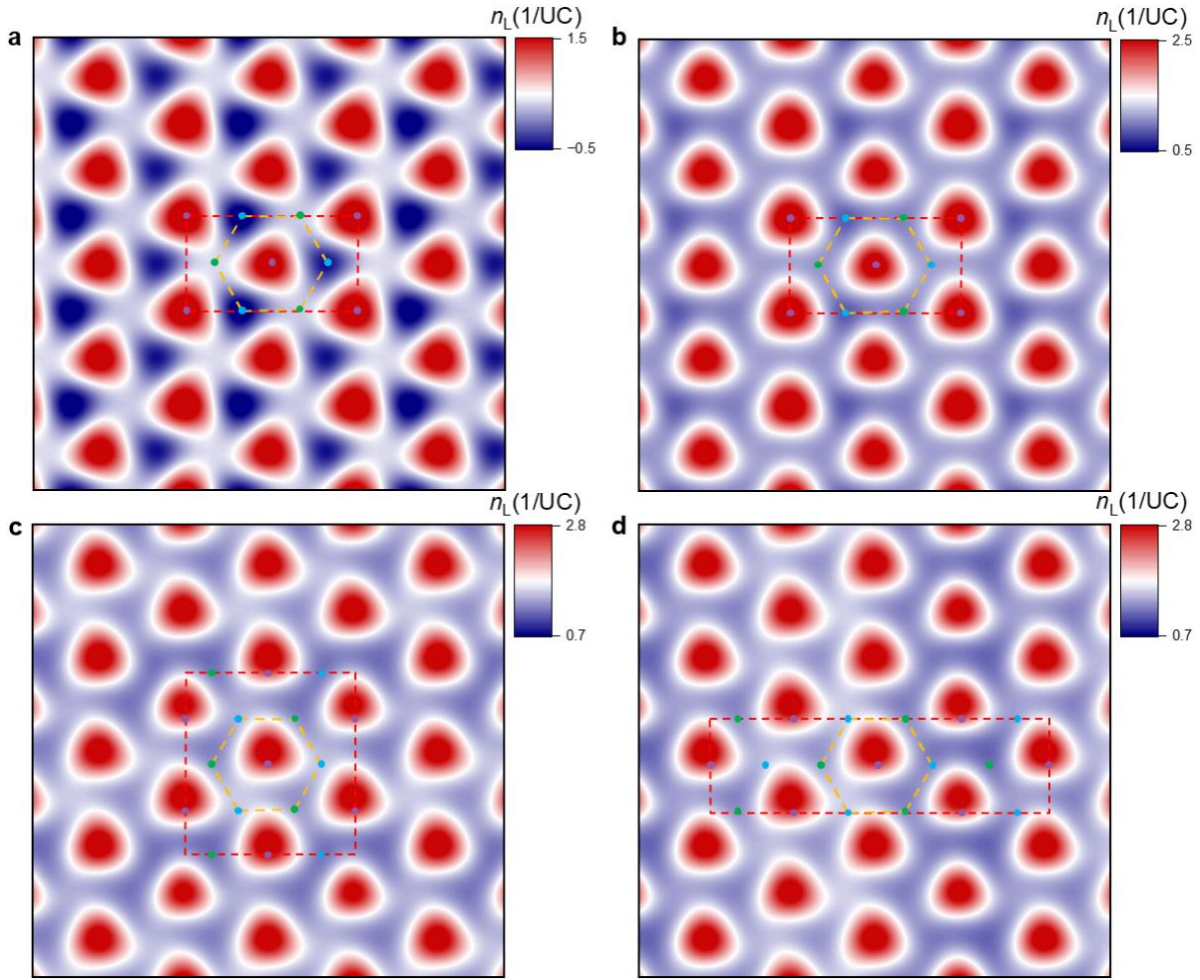

**Fig.S17|Electron distribution for ABABBC stacking order. a-d,** Electron distribution by Hartree-Fock calculation in real space at $v$=0.5 **(a)**, $v$=1.5 **(b)**, and $v$=1.75 for bubble phase **(c)** and strip phase **(d)**, the value $n_L$ represents for the number of electrons per moiré unit cell. The circle dots represent for different stacking sites. Purple dots represent for ABABBC stacking sites, blue dots represent for ABBCAB stacking sites, green dots represent for BCABAB stacking sites. The orange hexagonal dashed line represents for the original moiré unit cell, the red rectangular dashed line represents for the expanded moiré unit cell formed by CDW.

Fig.18 shows the real space density profiles of the flat bands in magic angle twist bilayer graphene (MATBG), ABABBC-stacking CTTBG and ABABAB-stacking CTTBG. The density profiles of flat bands in CTTBG are more extended than those in MATBG. This means that nonlocal Coulomb interactions are more important in CTTBG, which can intrigue the existence of CCDW as inter-site repulsion disfavors electrons occupying neighbor sites.

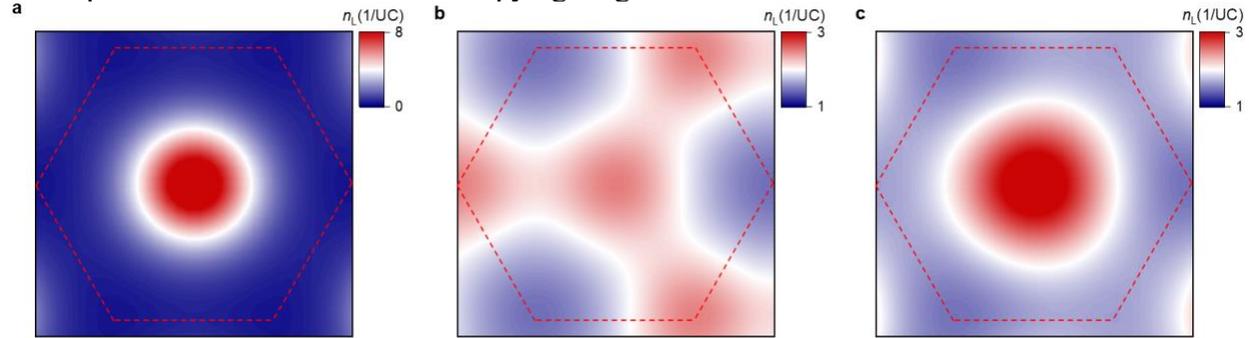

**Fig.S18|Comparison of real space density profiles of the flat bands in MATBG and CTTBG.**
**a-c,** the real space density profiles of the flat bands in MATBG **(a)**, ABABBC-stacking CTTBG **(b)** and ABABAB-stacking CTTBG **(c)**, the value $n_L$ represents for the number of electrons per moiré unit cell. The red dashed hexagons represent for the moiré unit cell.